\documentclass[amsmath,showpacs,aps,prb,reprint]{revtex4-1}
\usepackage{graphics}
\usepackage{graphicx}
\usepackage{amsmath}
\usepackage{amssymb}
\usepackage{enumerate}
\usepackage{xfrac}
\usepackage[caption=false]{subfig}
\usepackage[normalem]{ulem}
\usepackage{color}
\usepackage{txfonts}
\usepackage[mathscr]{euscript}
\usepackage[pdftex,
            pdfauthor={Pontus Laurell and Gregory A. Fiete},
            pdftitle={Magnon thermal Hall effect in kagome antiferromagnets with Dzyaloshinskii-Moriya interactions}]{hyperref}
\usepackage{verbatim}
\usepackage{gensymb}
\usepackage{ulem}
\newcommand\trick[1]{}	

\begin{document}
\title{Magnon thermal Hall effect in kagome antiferromagnets with Dzyaloshinskii-Moriya interactions}
\author{Pontus Laurell}
\email{laurell@physics.utexas.edu}
\author{Gregory A. Fiete}
\affiliation{Department of Physics, The University of Texas at Austin, Austin, TX 78712, USA}
\date{\today}

\begin{abstract}
We theoretically study magnetic and topological properties of antiferromagnetic kagome spin systems in the presence of both in- and out-of-plane Dzyaloshinskii-Moriya interactions. In materials such as the iron jarosites, the in-plane interactions stabilize a canted noncollinear ``umbrella'' magnetic configuration with finite scalar spin chirality. We derive expressions for the canting angle, and use the resulting order as a starting point for a spin-wave analysis. We find topological magnon bands, characterized by non-zero Chern numbers. We calculate the magnon thermal Hall conductivity, and propose the iron jarosites as a promising candidate system for observing the magnon thermal Hall effect in a noncollinear spin configuration. We also show that the thermal conductivity can be tuned by varying an applied magnetic field, or the in-plane Dzyaloshinskii-Moriya strength. In contrast with previous studies of topological magnon bands, the effect is found to be large even in the limit of small canting.
\end{abstract}
\maketitle

\section{Introduction}
In recent years there has been a growing interest in correlated materials with strong spin-orbit coupling (SOC).\cite{Witczak-Krempa2013,Rau2015,0034-4885-79-9-094504} When both electron-electron interactions and SOC are present novel phases may be found, including spin liquids,\cite{RevModPhys.88.041002,Savary2017} unconventional magnetic orders,\cite{PhysRevB.86.205124,PhysRevB.86.155127,PhysRevLett.112.077204} and fractional topological states.\cite{doi:10.1146/annurev-conmatphys-031115-011559,Maciejko:np15} One important class of such materials is Mott insulators at half-filling, which can be described using local moment (spin) models. When SOC is present it often gives rise to Dzyaloshinskii-Moriya interactions (DMI),\cite{PhysRevLett.4.228,PhysRev.120.91} which provides one route towards topologically nontrivial magnetic excitations in both ordered,\cite{PhysRevLett.104.066403,Onose2010,PhysRevB.85.134411,Hirschberger2015,PhysRevB.95.195137} and disordered\cite{PhysRevLett.104.066403,Hirschberger2015a,PhysRevB.91.125413} systems. Materials with these properties could potentially be exploited in antiferromagnetic spintronics,\cite{RevModPhys.90.015005} and topological magnonic devices.\cite{PhysRevApplied.9.024029}

As a result there have been many intriguing proposals to use magnetic excitations as a platform for realizing analogs of other topological systems. Recent examples include analogs of the Haldane-Kane-Mele model,\cite{PhysRevLett.117.227201} Dirac semimetals,\cite{Owerre2017b,PhysRevLett.119.247202} Weyl semimetals,\cite{Li2016,PhysRevB.95.224403} triple points,\cite{Hwang2017} and chiral topological insulators.\cite{Li2017} The most studied class of systems, however, is ordered magnetic insulators with topological magnon bands, associated with nonzero Chern numbers. These systems can be seen as bosonic, charge-neutral analogs of integer Chern insulators.\cite{PhysRevB.87.144101,PhysRevB.89.134409} An observable signature of the topology is found in the thermal Hall effect of magnons (or magnon thermal Hall effect), in which the magnon edge current produces a thermal Hall current and transverse thermal conductivity $\kappa_{xy}$ in the presence of a thermal gradient.\cite{PhysRevLett.104.066403,Onose2010,PhysRevLett.106.197202,PhysRevB.84.184406,doi:10.7566/JPSJ.86.011010} This effect has been experimentally observed in pyrochlore and kagome ferromagnets,\cite{Onose2010,PhysRevB.85.134411,Hirschberger2015,PhysRevLett.115.147201} where the DMI induces a non-trivial Berry phase on the magnons. A similar thermal Hall effect has also been observed above the ordering temperature in a kagome ferromagnet,\cite{Hirschberger2015} in the disordered phase of a spin ice material,\cite{Hirschberger2015a} and in a spin-liquid state.\cite{Watanabe2016} The existence of the signal in the disordered regime can be understood in terms of spin-linear response theory or Schwinger bosons.\cite{PhysRevB.91.125413,doi:10.7566/JPSJ.86.011007} Finally, a related phenomenon is the magnon spin Nernst effect, which can be seen as two copies of the magnon thermal Hall effect.\cite{PhysRevLett.117.217202,PhysRevLett.117.217203,PhysRevB.96.134425}

Despite the experimental emphasis on ferromagnetic systems, there is no principle ruling out the magnon thermal Hall effect in antiferromagnetically coupled systems.
\cite{PhysRevB.89.054420,PhysRevB.90.035114} Of particular interest is noncoplanar orders where a finite scalar spin chirality\cite{PhysRevLett.104.066403} $\mathbf{S}_i \cdot \left( \mathbf{S}_j \times \mathbf{S}_k\right)$ can produce the non-trivial topology.\cite{Owerre2016d} This is the reason why several theoretical predictions focus on intrinsically non-coplanar magnetic configurations such as in pyrochlore iridate thin films,\cite{PhysRevLett.118.177201} and bulk systems,\cite{Hwang2017}, or noncollinear configurations canted out-of-plane by a magnetic field. For the latter case, the magnon thermal Hall effect has been predicted on the star,\cite{Owerre2016d} honeycomb,\cite{0953-8984-28-38-386001} and kagome lattices.\cite{PhysRevB.90.035114,Owerre2016d,Owerre2017,Owerre2017c}

In this paper we study kagome quantum antiferromagnets with intrinsic non-coplanar orders, such as iron jarosites,\cite{PhysRevB.63.064430,PhysRevB.66.014422,PhysRevB.67.224435,PhysRevLett.96.247201,PhysRevB.73.214446,PhysRevB.83.214406,PhysRevB.85.094409} 
chromium jarosites,\cite{PhysRevB.56.8091,PhysRevB.64.054421} 
vesignieite,\cite{PhysRevB.88.144419} and the recently introduced Nd$_3$Sb$_3$Mg$_2$O$_{14}$ compound.\cite{PhysRevB.93.180407} Due to their intrinsic noncoplanarity several of these materials have been proposed as suited for experimental studies of the magnon thermal Hall effect.\cite{PhysRevB.95.014422,doi:10.7566/JPSJ.86.011007} However, their finite spin chirality (and thus topological properties) is believed to be due to in-plane components of the Dzyaloshinskii-Moriya (DM) vectors, which have been neglected in past works on the magnon thermal Hall effect in kagome systems. (They have, however, been studied in the related context of Weyl magnons in stacked kagome models.\cite{10.1139/cjp-2018-0059}) 
In systems with weak in-plane DMI this can be justified on energetic grounds,\cite{PhysRevB.78.140405} but it isn't necessarily weak in these materials. Furthermore, it has been shown that even weak in-plane DMI can have a significant impact on thermodynamic properties.\cite{PhysRevLett.98.207204,PhysRevB.76.184403} 
Our treatment includes the in-plane DMI, which we will show results in both higher Chern numbers, and a large magnon thermal Hall effect. The effect can be tuned by an applied transverse magnetic field, or the in-plane DM strength. Remarkably, we find a large effect even when the spin chirality is very small, which is in clear contrast to past works on noncollinear systems.

The paper is organized as follows. In Sec.II we introduce the spin model and derive new expressions for the canting angle, allowing for XXZ anisotropy, next-nearest neighbor exchange, and an applied magnetic field. We also comment on applicable materials. In Sec.III we carry out a spin-wave analysis about the groundstate. In Sec.IV the magnon thermal Hall conductivity is calculated, and shown to be tunable as a function of the applied field and in-plane DMI strength. Finally, in Sec.V we end the paper with conclusions.

\section{Model and magnetic order}
\subsection{Spin model}
\begin{figure}
      \begin{center}
		\includegraphics[width=.7\columnwidth]{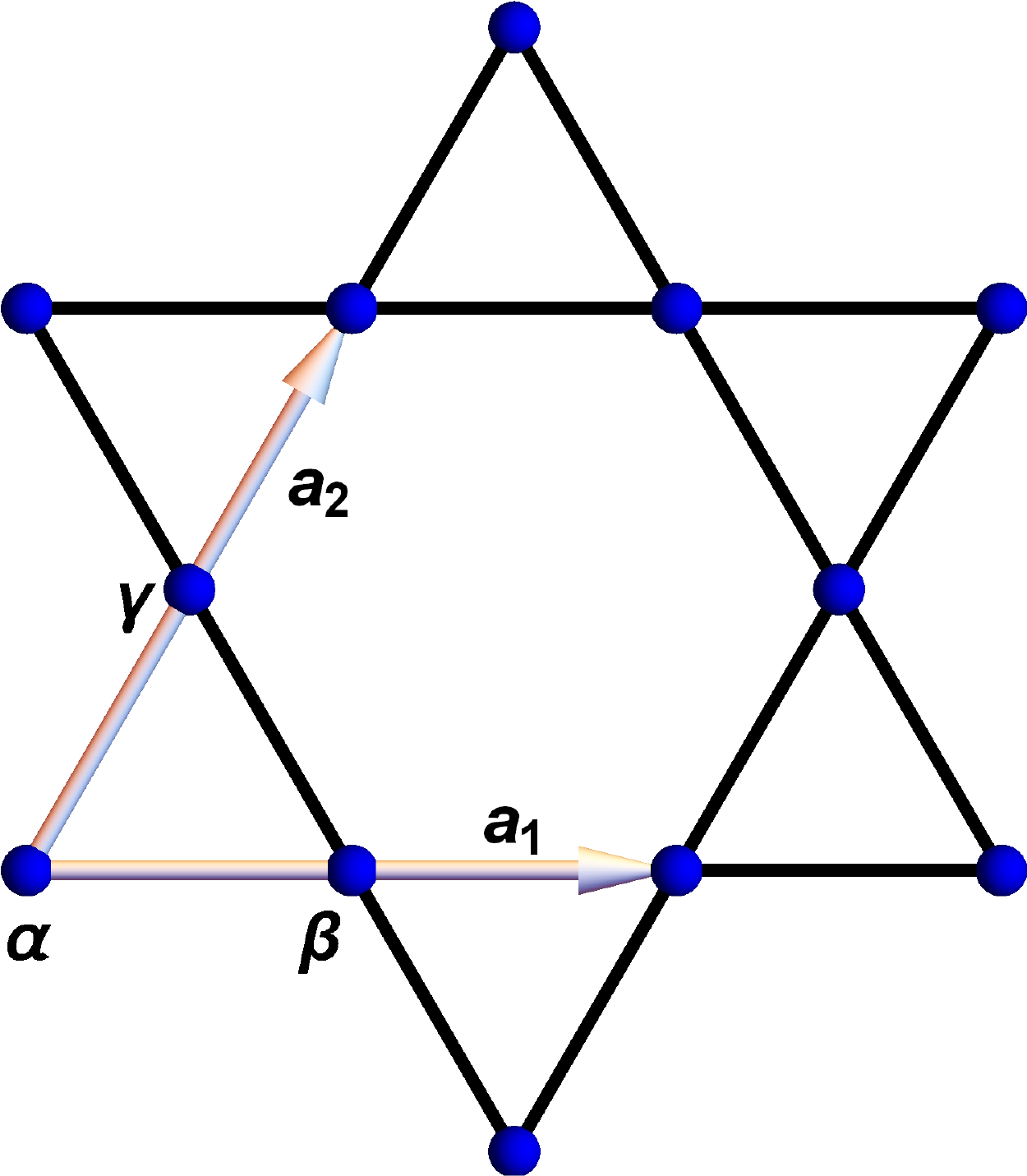}
		\caption{\label{kagomeLatticeVectors}The kagome lattice is a network of corner-sharing triangles. The two lattice vectors $\mathbf{a}_1$ and $\mathbf{a}_2$ are shown, along with the sublattices $\alpha,\beta,\gamma$.}
      \end{center}
\end{figure}

\begin{figure}
      \subfloat[][top]{
		\includegraphics[width=.8\columnwidth]{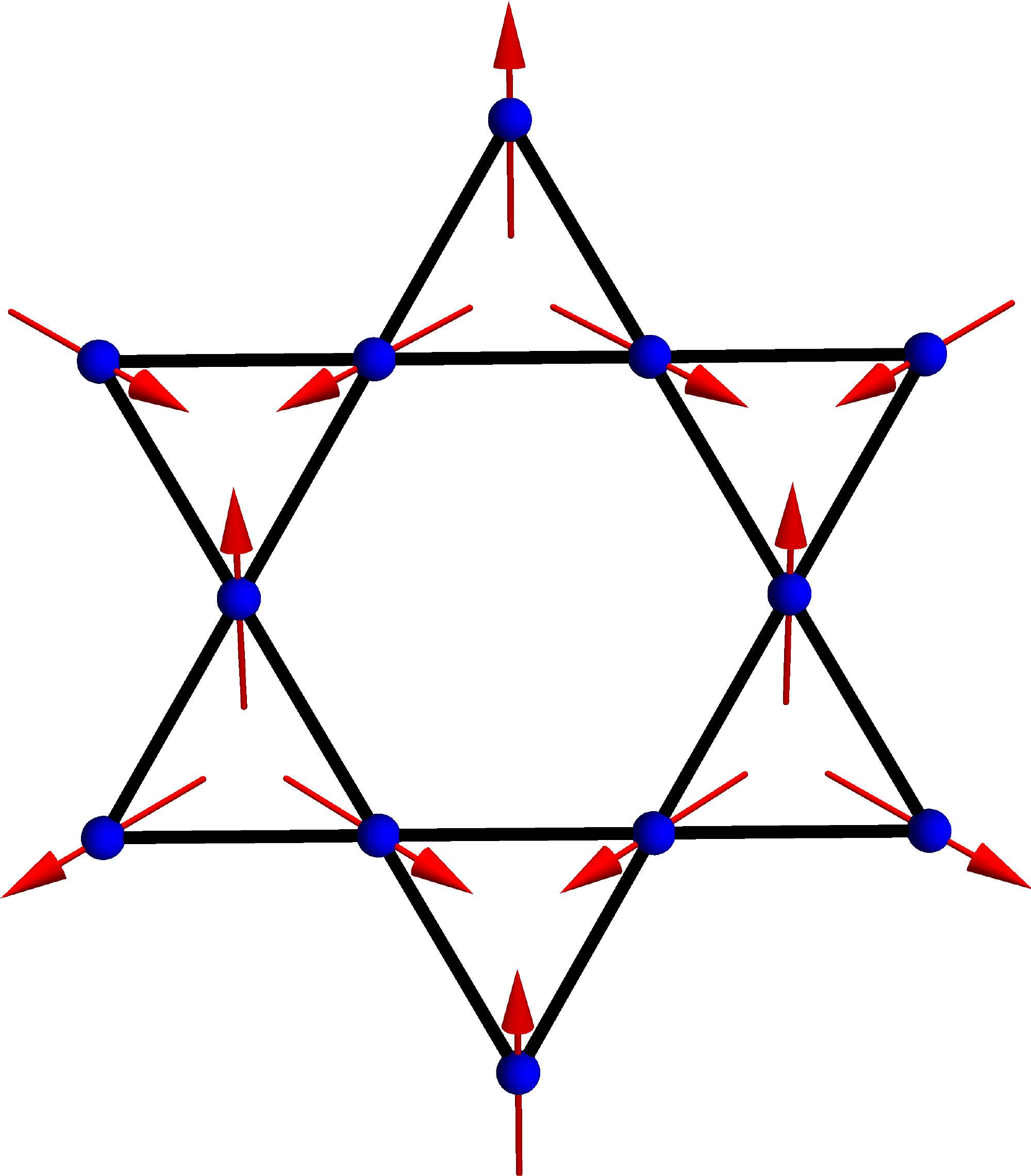}
	}\\
	\subfloat[][side]{
		\includegraphics[width=.8\columnwidth]{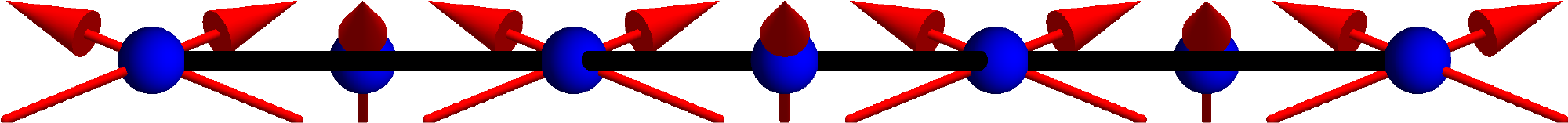}
	}
	\caption{\label{canted120fig}(a) Top, and (b) side views of the canted $120^\circ$ (or ``umbrella'') magnetic configuration. The top view also serves as illustration of the $120^\circ$ order without canting.}
\end{figure}

The kagome lattice is shown in Fig.~\ref{kagomeLatticeVectors}. It has the property that the middle point between two lattice sites is not an inversion center. Consequently, DMI is not forbidden by symmetry,\cite{PhysRevB.66.014422,PhysRev.120.91} and so a natural first spin model for many kagome Mott insulators is,
\begin{align}
      H_1	&=	\sum_{\langle i,j\rangle} \left[ J_1 \mathbf{S}_i\cdot \mathbf{S}_j + \mathbf{D}_{ij} \cdot \left( \mathbf{S}_i \times \mathbf{S}_j \right) \right],	\label{firstmodel}
\end{align}
where $J_1$ is the strength of the nearest-neighbor (NN) Heisenberg exchange, $\langle i,j\rangle$ denotes a sum over nearest neighbors, $\mathbf{D}_{ij}$ is the DM vector, and $\mathbf{S}_i$ is the vector of spin operators at site $i$. Since the kagome plane is a mirror plane, Moriya's rules prescribe that the DM vector points perpendicular to the plane, i.e. $\mathbf{D}_{ij} = D_z\hat{z}$.\cite{PhysRevB.66.014422,PhysRev.120.91} In the spin-$1/2$ case one expects the existence of a quantum critical point at $D_z^c/J_1 \simeq 0.1$ between a spin liquid for $|D_z|<D_z^c$ and a coplanar ordered state for $|D_z|>D_z^c$.\cite{PhysRevB.78.140405,PhysRevB.81.144432} For both spin-$1/2$ and higher spins, mean-field theory predicts two possible $\mathbf{q}=0$ states, with opposite sign chirality.\cite{PhysRevB.66.014422} For $D_z<0$ a state with all spins pointing towards or away from the center of triangles is selected, such as in Fig.~\ref{canted120fig}~(a). This state is known as the $120^\circ$ configuration. (Note that the DM vectors are determined by symmetry only up to a sign. Here we use the convention of Matan et al.,\cite{PhysRevB.83.214406} which is the opposite to that of Elhajal et al.\cite{PhysRevB.66.014422} In the latter convention the $120^\circ$ order would correspond to $D_z>0$.)

\begin{figure}
      \begin{center}
		\includegraphics[width=.7\columnwidth]{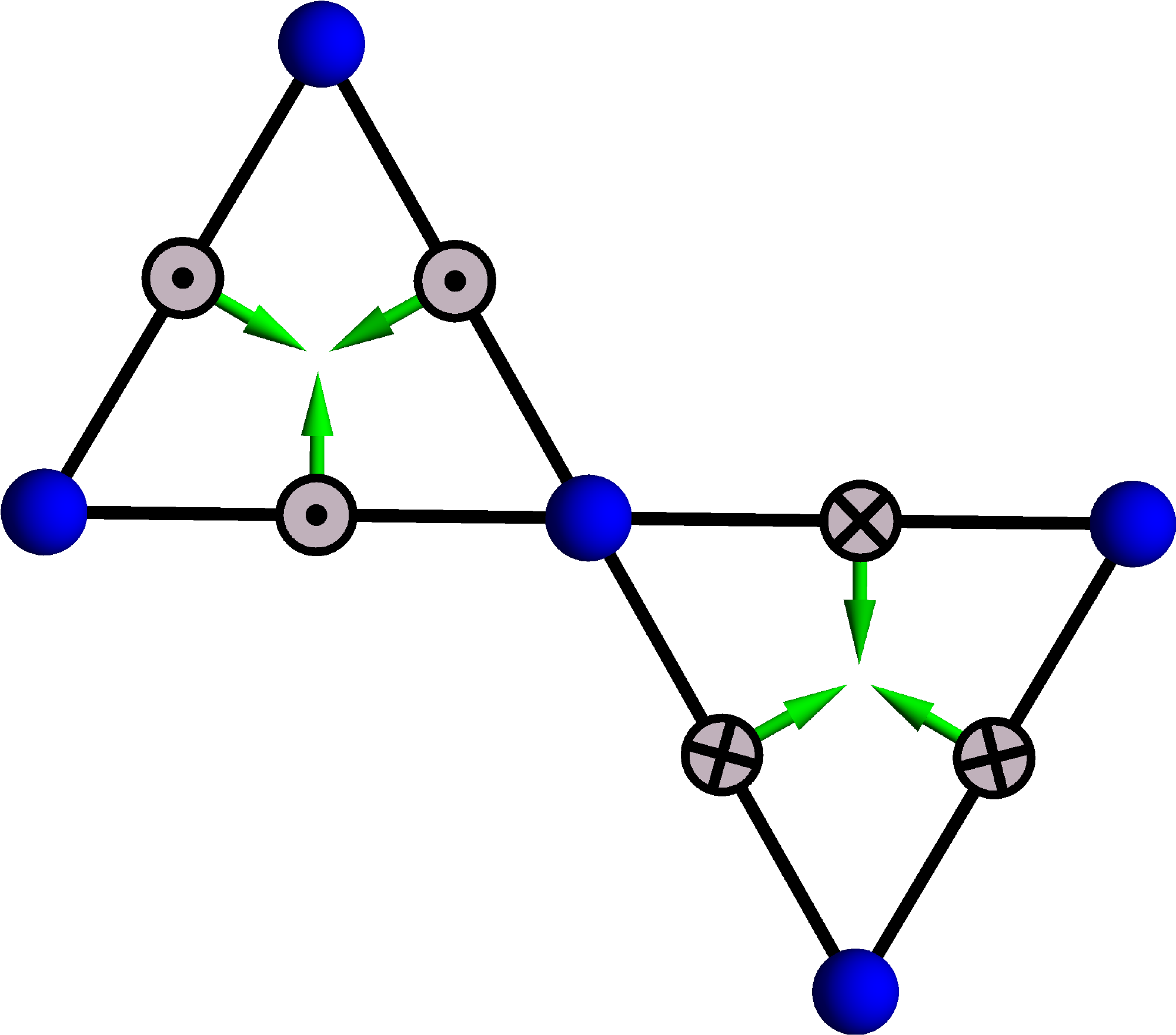}
		\caption{\label{kagomeDMVectors}(color online) The kagome DM vectors. The in-plane directions are shown by the green arrows, while the out-of-plane component is out-of (into) the plane for the triangles pointing up (down).}
      \end{center}
\end{figure}

If the symmetry is lowered such that the kagome plane no longer is a mirror plane, in-plane components of the DM vector are also allowed. This can occur in e.g. three-dimensional materials containing stacked kagome layers, such as jarosites, due to the local crystal environment near the kagome plane.\cite{PhysRevB.66.014422} In this case we can write $\mathbf{D}_{ij} = D_p \hat{n}_{ij} + D_z \hat{z}$, where $\hat{n}_{ij}$ is some in-plane DM (unit) vector, and $D_p$ is the strength of the in-plane DMI. For bond $(\alpha\beta)$ in Fig.~\ref{kagomeLatticeVectors} we write $\mathbf{D}_{\alpha\beta}=(0,D_p,D_z)$. Other DM vectors can then be obtained by rotation, and are shown in Fig.~\ref{kagomeDMVectors}. The in-plane DMI $D_p$ cants the spins out-of-plane, leading to a weak ferromagnetic moment. This is the so-called ``umbrella'' configuration shown in Fig.~\ref{canted120fig}~(b). Experimental results on the spin-$1/2$ material vesignieite BaCu$_3$V$_2$O$_8$(OH)$_2$ suggests that non-zero $D_p$ can stabilize this canted order when $|D_z|<D_z^c$.\cite{PhysRevB.88.144419} For the spin-$5/2$ iron jarosites one expects smaller frustration tendencies, and also finds this canted order.\cite{PhysRevLett.96.247201}

The model \eqref{firstmodel} is, however, not complete. In the iron jarosites there is also next-nearest neighbor (NNN) Heisenberg exchange, for example. For this reason we introduce an expanded model,
\begin{align}
      H	&=	H_{J_1} + H_{DM} + H_{J_2} + H_B,	\label{modelfull}
\end{align}
where
\begin{align}
      H_{J_1}	&=	J_1 \sum_{\langle i,j\rangle} \left( \mathbf{S}_i\cdot \mathbf{S}_j - \Delta S_i^z S_j^z \right),	\\
      H_{J_2}	&=	J_2 \sum_{\langle\langle i,j\rangle\rangle} \mathbf{S}_i \cdot \mathbf{S}_j,	\\
      H_{DM}	&=	\sum_{\langle i,j\rangle} \mathbf{D}_{ij} \cdot \left( \mathbf{S}_i \times \mathbf{S}_j \right),	\\
      H_B	&=	-B\sum_i S_i^z.
\end{align}
Above, $\Delta$ is an XXZ anisotropy for the NN exchange, $J_2$ is the NNN exchange strength, and $B$ is a magnetic field applied perpendicular to the kagome plane. For convenience we also define $J\equiv J_1 -J_1\Delta + J_2$. We neglect any interlayer coupling, because it is weak compared to the DMI in iron jarosites.\cite{PhysRevB.83.214406} The canted $120^\circ$ order can alternatively be explained by single-ion anisotropy terms,\cite{PhysRevLett.96.247201,0953-8984-18-39-015,PhysRevB.79.045102} but these are expected to be small for the Fe$^{3+}$ ions of iron jarosites.\cite{PhysRevB.66.014422} {We introduce the XXZ anisotropy to our model to help explain large canting angles in certain materials. It is, however, not necessary for the description of current experimental data on jarosites, and will be assumed to be zero in all calculations for jarosites.}

\subsection{Canting angle}
Neglecting quantum effects and fluctuations, the canted $120^\circ$ order can be described using classical spin vectors
\begin{align}
      \mathbf{S}_\alpha	&=	S\left( -\frac{\sqrt{3}}{2}\cos \eta, \, -\frac{1}{2}\cos \eta, \, \sin \eta \right)	,\\
      \mathbf{S}_\beta	&=	S\left( +\frac{\sqrt{3}}{2}\cos \eta, \, -\frac{1}{2}\cos \eta, \, \sin \eta \right)	,\\
      \mathbf{S}_\gamma	&=	S\left( 0, \, \cos \eta, \, \sin \eta \right)	,
\end{align}
where $S$ is length of the spin, and $\eta$ is the canting angle, with $\eta=0$ corresponding to the uncanted, coplanar order. The order leads to a scalar spin chirality in the unit cell
\begin{align}
      \chi_{\triangle}	&=	\mathbf{S}_\alpha \cdot \left(\mathbf{S}_\beta \times \mathbf{S}_\gamma \right) = \frac{3\sqrt{3}S^3}{2} \cos^2 \left( \eta\right) \sin \eta.	\label{scalarChiralityUmbrella}
\end{align}
The vector spin chirality is
\begin{align}
      \vec{\chi}_\triangle	&=	\mathbf{S}_\alpha \times \mathbf{S}_\beta + \mathbf{S}_\beta \times \mathbf{S}_\gamma + \mathbf{S}_\gamma \times \mathbf{S}_\alpha = \frac{3\sqrt{3}S^2}{2} \cos^2 \left( \eta\right)\hat{z}	\label{vectorChiralityUmbrella}
\end{align}
in the ordered state, but can remain non-zero even above the magnetic ordering temperature $T_N$.\cite{10.1038/nmat1353}

The classical energy for the Hamiltonian \eqref{modelfull} is then found to be
\begin{align}
      \frac{E^{cl} \left( \eta\right) }{NS^2}	&=	\frac{J}{2}-\frac{3J}{2}\cos 2\eta - 2J_1\Delta\sin^2 \eta - \frac{B}{S}\sin \eta\nonumber\\
				&+ \sqrt{3}D_z \cos^2 \eta + \sqrt{3}D_p\sin 2\eta.
\end{align}
The canting angle $\eta$ is determined by the zeros of
\begin{align}
      f\left( \eta \right)	&=	\frac{1}{NS^2} \frac{\partial E^{cl} \left( \eta\right)}{\partial \eta}.		\label{cantingDef}
\end{align}

To first order in $\eta$ the solution takes the intuitive form
\begin{align}
      \eta^{(1)}	\equiv \frac{B_\mathrm{eff}}{B_\mathrm{sat}},	
\end{align}
where the effective field and saturation field are given by
\begin{align}
      B_\mathrm{eff}	&=	\frac{B}{S} + 2\sqrt{3} \left| D_p\right|,	\label{cantingBeff}\\
      B_\mathrm{sat}	&=	6\left( J_1 + J_2\right) -4J_1 \Delta +2\sqrt{3}\left| D_z\right|.
\end{align}
Thus, both in-plane DMI and transverse magnetic fields can cause canting. The XXZ anisotropy increases the canting angle by lowering the saturation field, but does not cause continuous canting in the absence of in-plane DMI or fields.

The equation $f(\eta)=0$ can also be solved for special cases. For $D_p\neq 0$, $B=\Delta=0$ we find
\begin{align}
      \tan 2\eta	&=	\frac{-2D_p}{\sqrt{3}\left(J_1+J_2\right) -D_z},\label{cantingTan}
\end{align}
which is just an extended version of the formula derived by Elhajal et al.\cite{PhysRevB.66.014422}, also including the NNN exchange. For small angles $\eta$, $\tan 2\eta \approx \sin 2\eta$, which results in the formula used in Refs.~[\onlinecite{PhysRevB.73.214446,PhysRevB.83.214406}]. In the special case of $D_p=\Delta=0$, $B\neq 0$,
\begin{align}
      \sin \eta		&=	\frac{B/S}{6\left( J_1 + J_2\right) -2\sqrt{3}D_z},
\end{align}
consistent with the result of Owerre.\cite{PhysRevB.95.014422}

\subsection{Overview of experimentally relevant materials}
\begin{center}
\begin{table*}[t]
\caption{\label{matTable}Materials that order in the canted $120^\circ$ configuration with experimentally derived parameters.}
\begin{ruledtabular}
\begin{tabular}{llcccccccc}
 Material & & Ref. & $S$ & $T_N$ (K) & $\eta$  & $J_1$~(meV) & $J_2$~(meV) & $|D_p|/J_1$ & $D_z/J_1$\\\hline
 Potassium iron jarosite & KFe$_3$(OH)$_6$(SO$_4$)$_2$  & \cite{PhysRevLett.96.247201} & $5/2$ & 65 & $1.9^\circ$ & 3.18 & 0.11 & 0.062 & -0.062 \\
 Potassium iron jarosite (alternate fit) &   & \cite{PhysRevB.73.214446} &  &  &  & 3.225 & 0.11 & 0.068 & -0.06 \\
 Silver iron jarosite 1 & AgFe$_3$(OH)$_6$(SO$_4$)$_2$ & \cite{PhysRevB.83.214406} & $5/2$ & 59 & $1.8^\circ$ & 3.18 & 0.11 & 0.057 & -0.053 \\
 Silver iron jarosite 2 & AgFe$_3$(OD)$_6$(SO$_4$)$_2$ & \cite{PhysRevB.83.214406} & $5/2$ & 59 & $2.4^\circ$ & 3.18 & 0.11 & 0.075 & -0.053 \\
 Vesignieite & BaCu$_3$V$_2$O$_8$(OH)$_2$ & \cite{PhysRevB.88.144419} & $1/2$ & 9 & $6^\circ$ & 4.6 & N/A & 0.19 & -0.07 \\
 & Nd$_3$Sb$_3$Mg$_2$O$_{14}$ & \cite{PhysRevB.93.180407} & $1/2$ & 0.56 & $30.6^\circ$ & N/A & N/A & 0.8 & N/A \\
\end{tabular}
\end{ruledtabular}
\end{table*}\end{center}

In Table~\ref{matTable} we list materials known to order in the umbrella configuration, along with experimentally determined values for the ordering temperature, canting angle, and interaction strengths. The chromium jarosite KCr$_3$(OD)$_6$(SO$_4$)$_2$ with $S=3/2$, $T_N=4$K also orders in a $\mathbf{q}=0$ structure, but is reported to only have slight canting.\cite{PhysRevB.56.8091,PhysRevB.64.054421} The different iron jarosites have essentially similar behavior, so for the purposes of this paper we will use potassium iron jarosite as our model system. It does have the highest ordering temperature, which tends to produce a stronger magnon thermal Hall effect, as higher energy bands can be populated before the order breaks down from thermal fluctuations. It also has the advantages of an undistorted kagome lattice, and that large single crystals can be grown.\cite{10.1038/nmat1353}

We again stress the importance of the in-plane DMI for ordering by noting that vesignieite and the putative spin liquid herbertsmithite ZnCu$_3$(OH)$_6$Cl$_2$ have very similar out-of-plane DM strengths. The reported values for herbertsmithite is $|D_z|/J \approx 0.08$, and $|D_p|/J\approx 0.01$.\cite{PhysRevLett.101.026405} Finally, we would also like to comment on the interesting Nd$_3$Sb$_3$Mg$_2$O$_{14}$ compound, which has a considerable canting angle of $\eta=30.6\degree$. Scheie et al. estimated that $|D_p|/J_1\sim 0.8$ would be required to produce this angle, even in the limit of $D_z\rightarrow 0$.\cite{PhysRevB.93.180407} Their estimation used the sine approximation for Eq.\eqref{cantingTan}, which actually underestimates the value of $|D_p|/J$ required. Using Eq.\eqref{cantingTan} we instead find $|D_p|/J>1.5$. Although even stronger DMI have been predicted for some materials in first-principles calculations,\cite{Yadav2017} we speculate that, e.g., an XXZ anisotropy term could be present. If so it would contribute to the large observed canting angle, and allow for weaker $|D_p|/J$.

\section{Spin-wave analysis}
We now turn to a linear spin-wave analysis for potassium iron jarosite in the ordered regime ($T<T_N$). We start from the spin Hamiltonian \eqref{modelfull}, with $\Delta=0$. 
To describe deviations about a noncollinear ground state, we rotate the spin quantization axis at each site $i$ such that the $\tilde{z_i}$ axis in the new coordinate system points along the direction of the local moment $\langle \mathbf{S}_i\rangle$. This is achieved by the sublattice-dependent SO(3) rotation
\begin{align}
    S_i^a	&= \left[ R_i \left( \tilde{S}_i \right) \right]^a = R_i^{ab} \tilde{S}_i^b,	\label{sublatticeSpinRotation}
\end{align}
where $S_i^a$ is the spin operator with a global $z$ axis and $\tilde{S}_i^a$ has the axis unique to site $i$. Deviations from the groundstate are represented by Holstein-Primakoff bosons,\cite{Auerbach1994}
$\tilde{S}_i^z	=	s-a_i^\dagger a_i$,
\begin{align}
    \tilde{S}_i^+	=	\sqrt{2s-a_i^\dagger a_i} a_i,			\quad     &\tilde{S}_i^-	=	a_i^\dagger \sqrt{2s-a_i^\dagger a_i}.
\end{align}

The Hamiltonian is truncated to quadratic order, Fourier transformed, and written in a matrix form,
\begin{align}
	&H_\mathbf{k}	=	\mathbf{X}^\dagger_\mathbf{k} h\left( \mathbf{k}\right) \mathbf{X}_\mathbf{k},	\quad h \left( \mathbf{k} \right)	=	\left[ \begin{array}{cc} A \left( \mathbf{k} \right)	& B\left( \mathbf{k} \right) \\ B^\star \left(-\mathbf{k}\right)	& A^\star \left( -\mathbf{k} \right)\end{array}\right],\\
	&\mathbf{X}_\mathbf{k}	=	\left( a_\alpha\left(\mathbf{k}\right), a_\beta\left(\mathbf{k}\right),	a_\gamma\left(\mathbf{k}\right),  a^\dagger_\alpha\left(-\mathbf{k}\right), a^\dagger_\beta\left(-\mathbf{k}\right), a^\dagger_\gamma\left(-\mathbf{k}\right) \right)^T.\nonumber	\label{matrixHamiltonian}
\end{align}
{For more details, see Appendix A.} Since the system is bosonic, it has to be diagonalized paraunitarily to ensure that the Bogoliubov transformation preserves the bosonic commutation relations.\cite{Colpa1978} Thus we diagonalize $g h\left( \mathbf{k}\right)$, where
\begin{align}
	g	&=	\left( \begin{array}{cc}I_{3\times 3}	&	0\\	0	& -I_{3\times 3}\end{array}\right),
\end{align}
and $I_{3\times 3}$ is the $3\times 3$ identity matrix. We keep the physical states $|u_n\rangle$, with positive eigenvalues of $gh$, where $n$ is a band index.

\begin{figure}[t]
	    \subfloat[][Full spectrum]{	
		\includegraphics[width=\columnwidth]{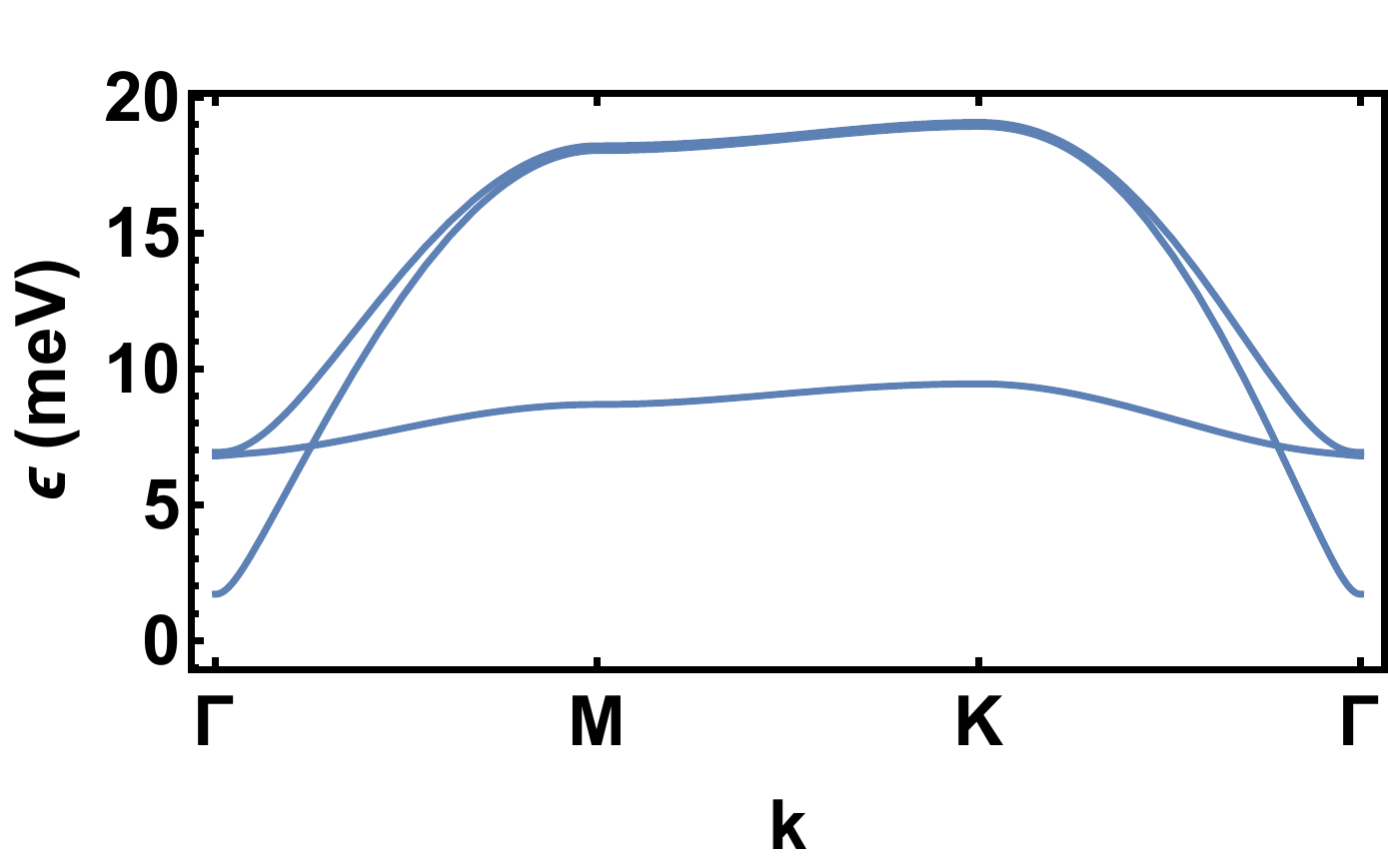}
	    }\\
	    \subfloat[][Zoomed in dispersion near $\Gamma$ and $K$ points]{
		\includegraphics[width=\columnwidth]{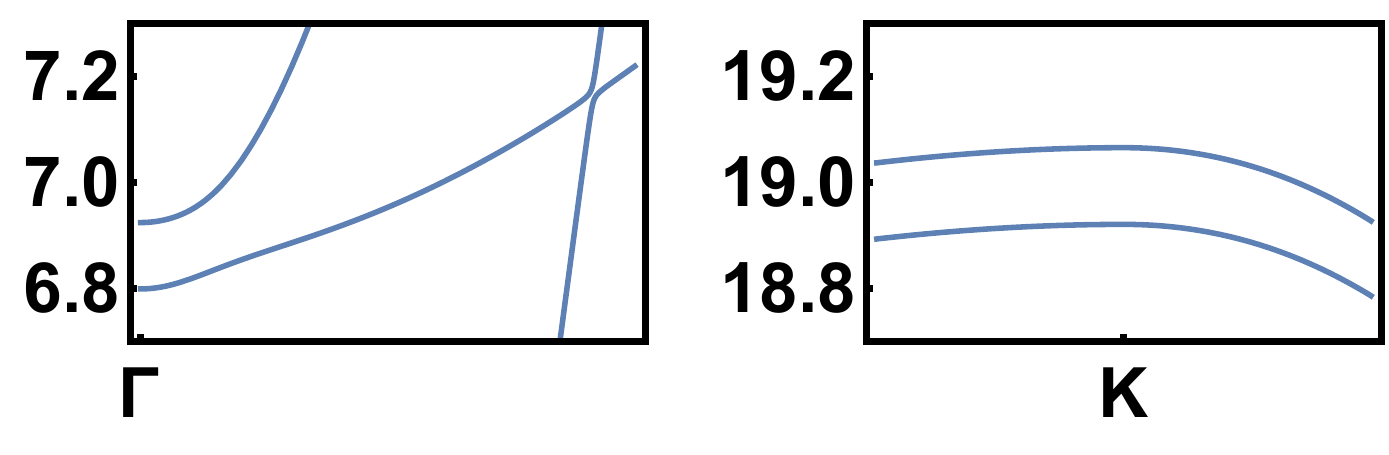}
	    }
      \begin{center}
	    \caption{\label{spectrum}Spin-wave spectrum for potassium iron jarosite. The full spectrum is shown in (a), and detailed views showing gaps and avoided crossings near the $\Gamma$ and $K$ points are shown in (b).}
      \end{center}
\end{figure}

Using the parameters for potassium iron jarosite from Matan et al.,\cite{PhysRevLett.96.247201} reproduced in Table~\ref{matTable}, we obtain the spin-wave spectrum shown in Fig.~\ref{spectrum} (a). {In Fig.~\ref{spectrum} (b) we provide close-up views of the dispersion near the $\Gamma$ and $K$ points, which show that the bands are separated by small gaps for these parameters, with a narrow avoided crossing between the lower two bands. The gaps between bands at the $\Gamma$, $M$, and $K$ symmetry points are $1.45$K, $2$K, and $3.46$K, respectively, in good agreement with previously obtained approximate analytical expressions.\cite{PhysRevB.73.214446} The bands remain separated throughout the entire Brillouin zone. This situation should be contrasted with the coplanar order for $D_p=0$, where there is a twofold protected degeneracy at the $\Gamma$ and $K$ points. The umbrella order, however, is associated with a trivial symmetry group,\cite{PhysRevB.83.184401} and hence has no protected degeneracies in the spin-wave spectra. Any gap closings would thus be accidental.
}

In the absence of NNN exchange there would be a very flat band throughout the entire Brillouin zone.\cite{PhysRevB.73.214446} Chernyshev proposes another mechanism for the observed dispersion of this band for $J_2=0$, in which the flat band is broadened by decay from higher bands.\cite{PhysRevB.92.094409,PhysRevB.92.144415} We also note that magnon-magnon interactions can have significant damping effects on topological magnon bands, as demonstrated in $S=1/2$ kagome ferromagnets.\cite{PhysRevLett.117.187203} 
For the purposes of our calculations, only the energetics of the dispersion is important, not its exact origin, so we proceed with the non-interacting phenomenological model with $J_2\neq 0$.

The presence of gaps allow us to define the Berry curvature $\Omega_n^z \left( \mathbf{k}\right)=i\left\langle\frac{\partial u_n}{\partial \mathbf{k}} \middle| \times \middle| \frac{\partial u_n}{\partial \mathbf{k}} \right\rangle$ for the $n^{th}$ magnon band, and calculate Chern numbers of individual bands,
\begin{align}
      C_n	&\equiv \frac{1}{2\pi}\int_\mathrm{BZ} \Omega_n^z \left( \mathbf{k}\right) \, dk^2.
\end{align}
In practice we calculate these numbers using a momentum space lattice discretization with $2000\times 2000$ lattice points.\cite{Fukui2005} We find the Chern numbers to have the unusual structure $(-3,+1,+2)$, going from the lowest to the highest band. Most previous works considering topological magnon bands on kagome lattices have only found two topological bands, with Chern number structures that are permutations of $(0,-1,+1)$.\cite{PhysRevB.89.134409,PhysRevB.95.014422,Owerre2017c,Seshadri2017,Owerre2018a,Owerre2018} The exception appears to be Mook et al.,\cite{PhysRevB.90.024412,PhysRevB.91.224411} who found several regions with higher Chern numbers for ferromagnetic $J_1$, $J_2$, and $D_p=0$. For the current case of antiferromagnetic $J_1$, the higher Chern numbers appear to be linked to the in-plane DMI $D_p$, as they can be found also when $J_2=0$. {Given that the bands are only separated by small gaps in this system, it is natural to wonder whether the Chern numbers are sufficiently protected to e.g. broadening effects. This question is, however, somewhat moot as the main experimental signature, the magnon thermal Hall conductivity, is due to the Berry curvature rather than the Chern number. (Since this is a bosonic system, the thermal Hall effect is not quantized.) The Berry curvature itself will persist until an actual gap closing.}

\section{Magnon thermal Hall effect}
\begin{figure}
      \begin{center}
	    \includegraphics[width=\columnwidth]{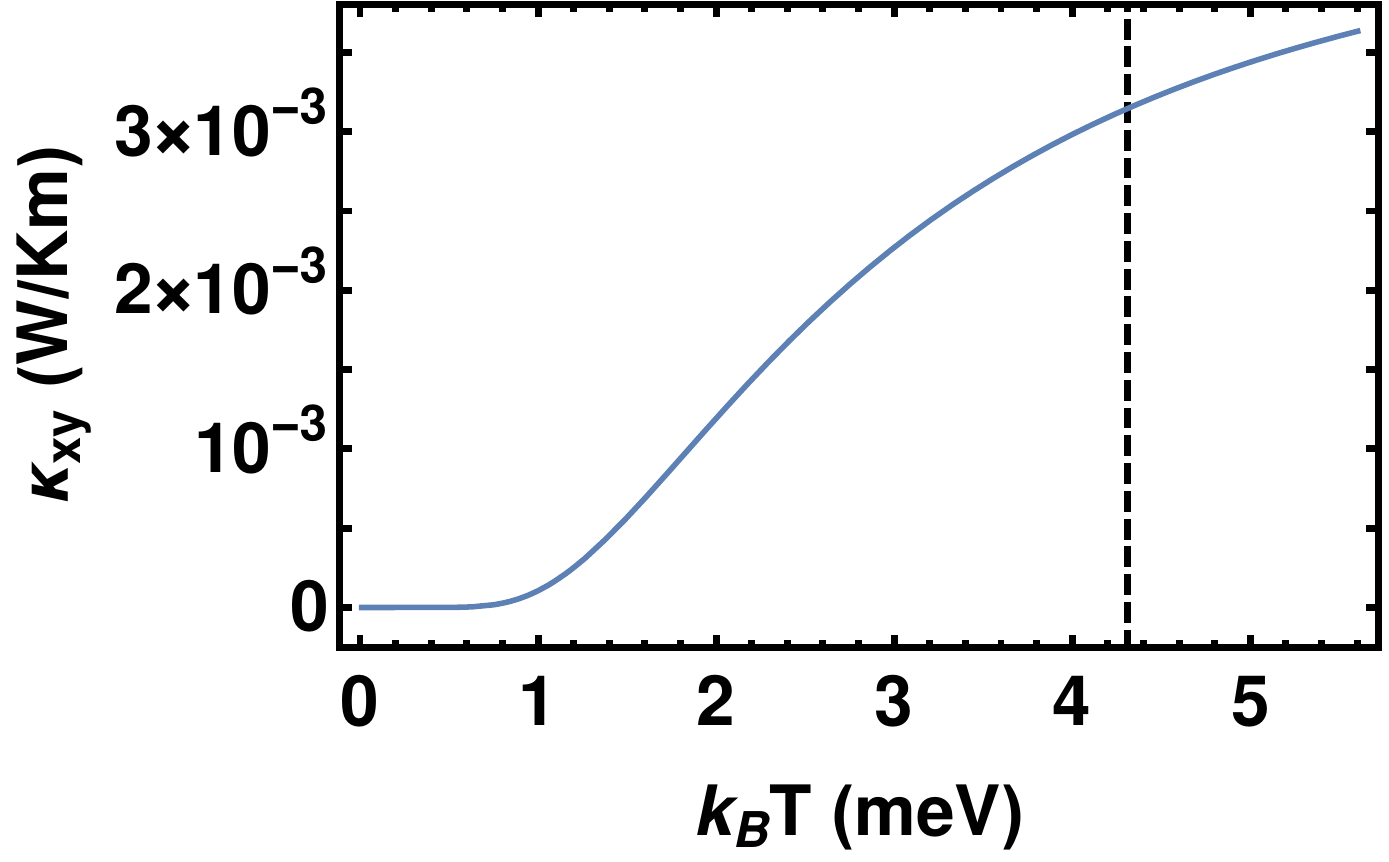}
	    \caption{\label{kappaT}Magnon thermal Hall conductivity for potassium iron jarosite in zero magnetic field as a function of temperature, up to the ordering temperature $T_N=65$K. The dashed vertical line marks $T_0=50$K.}
      \end{center}
\end{figure}
The magnon thermal Hall conductivity can be expressed
\begin{align}
      \kappa_{xy}	&=	-\frac{k_B^2 T}{\left( 2\pi\right)^2 \hbar} \sum_n \int_{BZ} \left[ c_2 \left[ g\left( \epsilon_{nk} \right) \right] - \frac{\pi^2}{3} \right] \Omega^z_{n} \left(\mathbf{k}\right) \, d^2k.	\label{kappaxy}
\end{align}
This formula is valid for a two-dimensional lattice and general spin-wave Hamiltonians that do not necessarily conserve magnon number.\cite{PhysRevB.89.054420,PhysRevB.89.134409}\textsuperscript{,}\footnote{The $\pi^2/3$ term vanishes upon integration over the full Brillouin zone since it is proportional to the sum of Chern numbers for all bands. However, the term can be important in cases when the integral is not taken over a closed surface.\cite{PhysRevB.89.054420}\protect\trick.} For stacks of two-dimensional layers, we can simply divide by the interlayer distance $l$ to find the thermal conductivity in units of W/Km.\footnote{Strictly speaking, one should generally use the distance on which the magnetic configuration repeats. In the case of the iron jarosites the magnetic unit cell includes six kagome layers, but the configuration shown in Fig.~\ref{canted120fig}, where spins point out of an upwards-pointing triangle, is repeated every two layers.\cite{PhysRevB.67.224435,PhysRevB.83.214406,10.1038/nmat1353} The layer in-between has the spins point into the upwards-pointing triangle. Since we neglect interlayer coupling, the exact alignment of sites is not required, and it is thus enough to consider two kagome planes. However, the net out-of-plane ferromagnetic moment also alternates in direction between layers.\cite{PhysRevB.83.214406,10.1038/nmat1353} When accounting for both the opposite sign canting angle and the reversed in-plane spin directions, it is found that all kagome planes contribute to the signal with the same sign and magnitude, allowing us to restrict our calculations to a single plane, and to use the interlayer distance. In contrast, e.g. Nd$_3$Sb$_3$Mg$_2$O$_{14}$ has the same sign net out-of-plane spin component in all layers.\cite{PhysRevB.93.180407}\protect\trick.} In Eq.~\eqref{kappaxy}, $g\left( \epsilon_{nk} \right)$ is the Bose-Einstein distribution, and the $c_2$ function is defined as
\begin{align}
	c_2 \left( x \right)		&=	\left( 1+x \right) \left[ \ln \left( \frac{1+x}{x} \right) \right]^2 - \left[ \ln x \right]^2 - 2\mathrm{Li}_2 \left( -x \right),
\end{align}
where $\mathrm{Li}_n \left( x \right)$ is the polylogarithm. It has the limits $c_2 \left[ g\left( \epsilon_{nk} \right) \right] \rightarrow \pi^2/3$ ($ 0 $) as $\beta \epsilon \rightarrow 0$ ($\beta \epsilon \rightarrow \infty$). Hence, $\kappa_{xy}$ vanishes as $T\rightarrow 0$ and is generically non-zero for higher temperatures.

We calculate $\kappa_{xy}$ for full magnetization up to $T_N=65$K, using the parameters for potassium iron jarosite of Matan et al.\cite{PhysRevLett.96.247201} listed in Table~\ref{matTable}. The conductivity is plotted as a function of temperature in Fig.~\ref{kappaT}, showing a monotonically rising behavior. Close to $T_N$ we expect a decrease in the signal due to decrease in sublattice magnetization. {Since the magnetic transition is experimentally observed to be mean-field like,\cite{10.1038/nmat1353,PhysRevB.83.214406} we will use the fixed magnetization approximation up $T_0=50$K${}\sim{}3T_N/4$ going forward. While linear, or non-interacting, spin-wave theory is most accurate at low temperatures for collinear systems, this is a noncollinear system with potentially important interaction effects. It thus remains to be seen whether the approximation is justified up to $T_0$ or not.} 
The signal at $T_0$ is $1.8\cdot 10^{-12}$ W/K for a single kagome layer. Using the interlayer distance $l \sim 5.7$\AA,\cite{PhysRevB.33.4919,PhysRevB.67.064401} we find a thermal conductivity of $3.15\cdot 10^{-3}$ W/Km at $T_0$. (If we instead use the DMI parameters fit by Yildirim and Harris,\cite{PhysRevB.73.214446} a slightly higher value of $3.41\cdot 10^{-3}$ W/Km at $T_0$ is found. The silver iron jarosites AgFe$_3$(OX)$_6$(SO$_4$)$_2$ yield similar values, $2.91\cdot 10^{-3}$~W/Km or $3.15\cdot 10^{-3}$~W/Km at $T=T_0^\mathrm{Ag}=45$K, for X=H and X=D, respectively.) The predicted values are on the same or better order as experimental observations for the magnon thermal Hall effect,\cite{Onose2010,PhysRevB.85.134411,Hirschberger2015,Hirschberger2015a} and predicted values for pyrochlore iridates in the noncollinear all-in--all-out spin configuration.\cite{Hwang2017}

\begin{figure}
      \begin{center}
	    \includegraphics[width=\columnwidth]{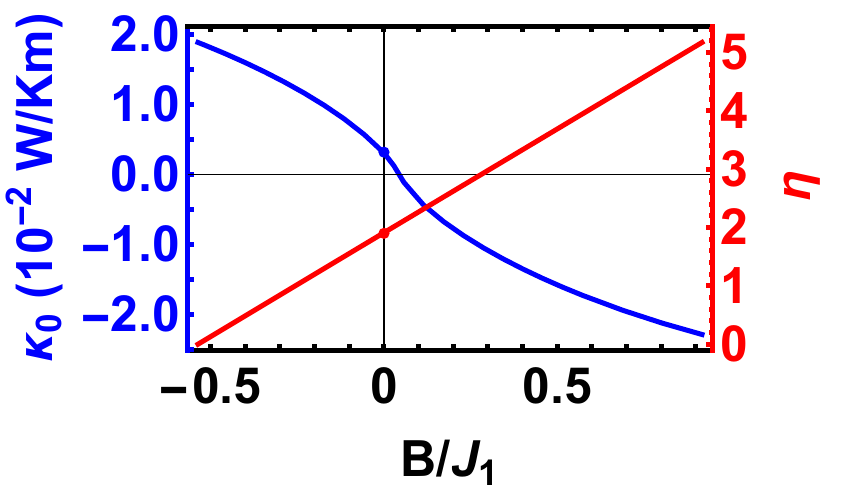}
	    \caption{\label{BScalingMatan}(color online) The magnon thermal Hall conductivity as a function of an applied staggered magnetic field perpendicular to the kagome plane. The magnon thermal Hall signal at $T=T_0$ is shown in blue, and the canting angle (in degrees) is shown in red. The sign change of the conductivity is associated with a sign reversal of all the Chern numbers. We note that the effect does not vanish with the scalar spin chirality as the canting angle approaches zero. The disks correspond to the values for iron jarosites in ambient conditions.}
	    \end{center}
\end{figure}

\begin{figure}
      \begin{center}
	    \includegraphics[width=\columnwidth]{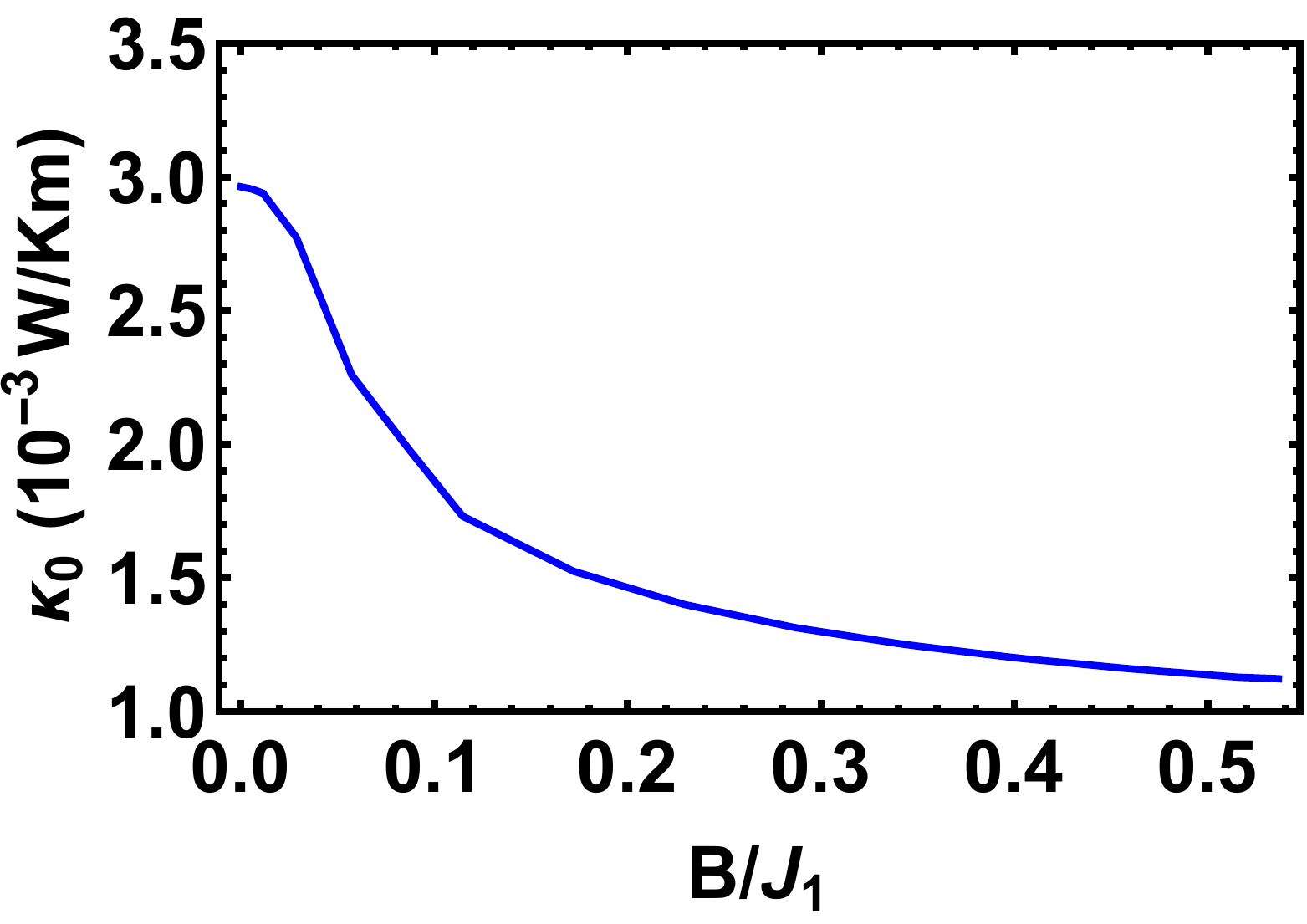}
	    \caption{\label{BScalingMatan2}(color online) The magnon thermal Hall conductivity as a function of an applied non-staggered magnetic field perpendicular to the kagome plane. The magnon thermal Hall signal at $T=T_0$ is shown in blue, and is obtained by averaging the responses for positive and negative fields in Fig.~\ref{BScalingMatan}.}
	    \end{center}
\end{figure}

We next explore regions of higher canting, where stronger effects are expected.\cite{PhysRevB.95.014422,Owerre2017,Owerre2016d} Recalling the effective field Eq.~\eqref{cantingBeff}, we achieve this by tuning the applied transverse field, or the in-plane DMI strength. In Fig.~\ref{BScalingMatan} we plot the size of the effect $\kappa_0$ at $T=T_0$ as a function of the applied field, along with the canting angle determined from Eq.~\eqref{cantingDef}. 
Since the sign of the out-of-plane spin components alternate in sign between adjacent kagome planes in potassium iron jarosite,\cite{PhysRevB.67.224435,PhysRevB.83.214406,10.1038/nmat1353} the values plotted in Fig.~\ref{BScalingMatan} are valid for a staggered field. If the field is not staggered, different layers will be canted differently, producing different conductivity values. As shown in Fig.~\ref{BScalingMatan2} there is, however, still a tunable signal. We also note that, in other materials, such as Nd$_3$Sb$_3$Mg$_2$O$_{14}$ the sign of this net ferromagnetic moment does not alternate,\cite{PhysRevB.93.180407} avoiding this complication. In such systems we expect a behavior similar to Fig.~\ref{BScalingMatan} in non-staggered fields.\footnote{There is also a field-induced transition in the iron and silver jarosites at a sufficiently strong ``non-staggered'' magnetic field, to an order where the sign of the net ferromagnetic moment does not alternate in sign. However, this occurs only above a rather high critical magnetic field strength $H_c \approx 14$T at $T=T_0$ for potassium iron jarosite,\cite{10.1038/nmat1353} and $H_c \approx 8$T at $T=T_0^\mathrm{Ag}$ for silver jarosite,\cite{PhysRevB.83.214406} above the field strengths we consider here.}

The range of staggered fields we consider in Fig.~\ref{BScalingMatan} is $B/J_1 \in [-.54, .92]$, or $B\in [-4.67, 8]$T using the effective magnetic moment $\mu_\mathrm{eff}=6.3\mu_B$.\cite{10.1038/nmat1353} The lower limit is chosen to make the canting angle approach zero. Note that we cannot reach $\eta = 0$ exactly, as the coplanar order would have protected degeneracies. Hence, the smallest canting angle plotted in Fig.~\ref{BScalingMatan} is $\eta \approx 1.12\times10^{-6}{}^\circ$, below which we run into numerical instabilities. As is expected, a positive (staggered) magnetic field (directed along $+\hat{z}$ in the first kagome plane) increases the canting angle, and affects the size of the effect. It decreases, and then shifts sign at $B/J_1 \approx 0.06$ due to a topological phase transition in which all Chern numbers change sign from the $(-3,+1,+2)$ structure to $(+3,-1,-2)$. {This transition may have additional structure, as discussed in Appendix B.} Following the transition, the effect gets enhanced by almost an order of magnitude at the largest fields. A field along $-\hat{z}$ decreases the canting angle, but remarkably also produces a stronger transport signature even as the canting angle approaches zero, and the gap at the $\Gamma$ point vanishes. This behavior is in stark contrast to previous results on kagome and star lattices that linked the signal in noncollinear systems directly to the scalar spin chirality,\cite{PhysRevB.95.014422,Owerre2017,Owerre2016d} and found that the effect vanished in the absence of canting. Our results underline the importance and subtle role of the in-plane DMI.
\begin{figure}
      \begin{center}
	  \includegraphics[width=\columnwidth]{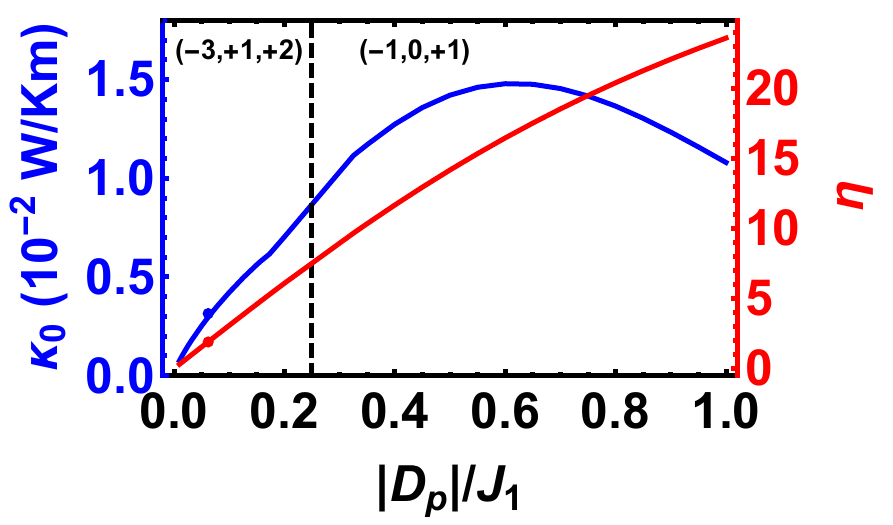}
	  \caption{\label{DpScalingMatan}(color online) The magnon thermal Hall conductivity at $T=T_0$ as a function of the in-plane component of the DM interaction is shown in blue. In red, the canting angle (in degrees) is plotted. Near $|D_p|/J_1=0.25$ there is a topological transition from a situation where all bands are topological, to one where the middle band is topologically trivial. The blue disk marks the value obtained in Fig.~\ref{kappaT} for the experimental canting angle $1.9^\circ$ (marked by a red disk).}
      \end{center}
\end{figure}

In Fig.~\ref{DpScalingMatan} we instead vary the strength of the in-plane DMI at zero magnetic field, while keeping the other parameters fixed. There is a topological transition near $|D_p|/J_1=0.25$ where one band becomes topologically trivial, and the more common $(-1,0,+1)$ structure is obtained. While the canting angle increases monotonically with $|D_p|$, the effect has a maximum at $|D_p|/J_1=0.6$. {The reason for this maximum is that the bands begin to separate, while also moving to higher energies for higher $|D_p|$. Hence, both the Berry flux and the occupation (at fixed temperature) are reduced. See Appendix B for the evolution of the spectrum as a function of $|D_p|$. 
However, both $B$ and $D_p$ can potentially stabilize the umbrella order at higher temperatures $T>T_N$, a fact we do not account for here. A correspondingly higher $T_0$ value would result in higher occupation numbers for large $D_p$, and larger effects throughout the parameter space considered. }
The effect vanishes as $|D_p|/J_1$ approaches zero, along with the spin chirality. Taken together, the results in Figs.~\ref{BScalingMatan} and \ref{DpScalingMatan} imply that the in-plane component of the DMI has important effects on the topology of the magnon bands, and on the magnon thermal Hall conductivity.

\section{Conclusion}
We have predicted a strong and tunable magnon thermal Hall effect in iron jarosites. These systems are also good insulators, which implies that there should be a clear transport signal. We thus propose that iron jarosites are a promising experimental candidate system to observe the effect for a noncollinear magnetic configuration. While we have focused on the topological edge transport, spin-wave transmission through the bulk is generically also present. However, it tends to get suppressed by disorder, which only weakly affects the topological edge modes.\cite{PhysRevB.97.081106} In addition to tuning by magnetic fields, additional control may be achieved by applying pressure, which can increase the ordering temperature in the jarosites.\cite{Coomer2006} 
More generally, topological transitions in magnon band structures may be driven by periodic driving,\cite{Owerre2018} and strain.\cite{Owerre2018a}

Our results show that 
in-plane Dzyaloshinskii-Moriya interaction can have important effects on the topology of the magnons, including higher Chern numbers. Most strikingly, we find that it can induce a magnon thermal Hall effect even when a staggered magnetic field is applied to make the canting angle and scalar spin chirality very small. In collinear systems, an out-of-plane DMI acts as an effective magnetic field on the magnons, and for noncollinear systems, a noncoplanar spin texture can induce the nontrivial topology. In the latter case, the magnon thermal Hall conductivity vanishes along with noncoplanarity. Our system, however, does not fall into either class. In fact, it is not entirely clear what mechanism produces the nontrivial topology in this limit, but the in-plane DMI must play a role.

\emph{Note added:} A recent experimental investigation [\onlinecite{PhysRevLett.121.107203}] of deuterated vesignieite finds another magnetic structure and spin Hamiltonian than assumed in this paper.

\section{Acknowledgements}
We gratefully acknowledge funding from ARO grant W911NF-14-1-0579, NSF DMR-1507621, and NSF MRSEC DMR-1720595. GAF gratefully acknowledges support from a Simons Fellowship. We acknowledge the Texas Advanced Computing Center (TACC) at The University of Texas at Austin for providing computing resources that have contributed to the research results reported within this paper. \url{www.tacc.utexas.edu}

\appendix
\renewcommand{\thefigure}{A\arabic{figure}}
\setcounter{figure}{0}

\section{Spin-wave Hamiltonian}
As described in the main text and Eq.~\eqref{sublatticeSpinRotation} we introduce sublattice-dependent transformations rotating the local $z_i$ axis onto the direction of the local magnetic moment. This can be implemented using the rotation matrices
\begin{align} R_\alpha &= \left(
\begin{array}{ccc}
 \frac{1}{4} (3 \sin (\eta )+1) & \frac{\sqrt{3}}{4} (\sin (\eta )-1) & -\frac{\sqrt{3}}{2} \cos (\eta ) \\
 \frac{\sqrt{3}}{4} (\sin (\eta )-1) & \frac{1}{4} (\sin (\eta )+3) & -\frac{\cos (\eta )}{2} \\
 \frac{\sqrt{3} }{2} \cos (\eta ) & \frac{\cos (\eta )}{2} & \sin (\eta ) \\
\end{array}
\right),
\end{align}
\begin{align}
  R_\beta &=\left(
\begin{array}{ccc}
 \frac{1}{4} (3 \sin (\eta )+1) & -\frac{\sqrt{3}}{4} (\sin (\eta )-1) & \frac{\sqrt{3}}{2} \cos (\eta ) \\
 -\frac{\sqrt{3}}{4} (\sin (\eta )-1) & \frac{1}{4} (\sin (\eta )+3) & -\frac{\cos (\eta )}{2} \\
 -\frac{\sqrt{3}}{2} \cos (\eta ) & \frac{\cos (\eta )}{2} & \sin (\eta ) \\
\end{array}
\right)
\end{align}
\begin{align}
	R_\gamma &= \left(
\begin{array}{ccc}
 1 & 0 & 0 \\
 0 & \sin (\eta ) & \cos (\eta ) \\
 0 & -\cos (\eta ) & \sin (\eta ) \\
\end{array}
\right).
\end{align}
We write the original spin Hamiltonian on the general form
\begin{align}
	H	&=	\sum_{ij} S_i^a \left(\Lambda_{ij}^{ab}+ \Xi_{ij}^{ab} \right) S_j^b,
\end{align}
where $\Lambda$ and $\Xi$ are the nearest (NN) and next-nearest neighbor (NNN) interaction matrices, respectively. After the spin rotation (a linear transformation) the interaction matrices are transformed into $\tilde{\Lambda}_{ij}^{ab} = \left[ R^T_i \Lambda_{ij} R_j\right]^{ab}$, $\tilde{\Xi}_{ij}^{ab} = \left[ R^T_i \Xi_{ij} R_j\right]^{ab}$.

The submatrices $A(\mathbf{k})$ and $B(\mathbf{k})$ of the spin-wave Hamiltonian Eq.~\eqref{matrixHamiltonian} are built from sums of elements of the rotated interaction matrices.  We first split them into nearest (NN) and next-nearest neighbor (NNN) terms,
\begin{align}
	A(\mathbf{k})	&=	A^\mathrm{NN}(\mathbf{k}) + A^\mathrm{NNN}(\mathbf{k}),\\
	B(\mathbf{k})	&=	B^\mathrm{NN}(\mathbf{k}) + B^\mathrm{NNN}(\mathbf{k}).
\end{align}
We have
\begin{align}
	A^\mathrm{NN}_{\mu\nu} \left( \mathbf{k}\right)	&=	S \left[ \tilde{\Lambda}_{\mu\nu}^{xx} + \tilde{\Lambda}_{\mu\nu}^{yy} -i\tilde{\Lambda}_{\mu\nu}^{xy} + i\tilde{\Lambda}_{\mu\nu}^{yx}\right] \cos \left( \mathbf{k}\cdot \mathbf{r}_{\mu\nu} \right)\nonumber\\
												&-S\delta_{\alpha\beta} \sum_\gamma \tilde{\Lambda}^{zz}_{\alpha\gamma},	\\
	B^\mathrm{NN}_{\mu\nu} \left( \mathbf{k}\right)	&=	S \left[ \tilde{\Lambda}_{\mu\nu}^{xx} - \tilde{\Lambda}_{\mu\nu}^{yy} +i\tilde{\Lambda}_{\mu\nu}^{xy} + i\tilde{\Lambda}_{\mu\nu}^{yx}\right] \cos \left( \mathbf{k}\cdot \mathbf{r}_{\mu\nu}\right),
\end{align}
where $\mu \in \{ \alpha,\beta,\gamma\}$, $\nu \in \{ \alpha,\beta,\gamma\}$ are sublattice indices. $\mathbf{r}_{\mu\nu}\equiv \mathbf{r}_\mu - \mathbf{r}_\nu$ are vectors connecting nearest neighbors. $A^\mathrm{NNN}(\mathbf{k})$ and $B^\mathrm{NNN}(\mathbf{k})$ are given by analogous expressions in terms of $\tilde{\Xi}$. Explicit expressions for the $A$ and $B$ matrices are not given here --- the number of cosine and sine terms introduced by the rotation would make their inclusion impractical.

\section{Band evolution as function of \texorpdfstring{$B$ and $D_p$}{B and Dp}}
\begin{figure}[tp!]
	\begin{center}
		\includegraphics[width=\columnwidth]{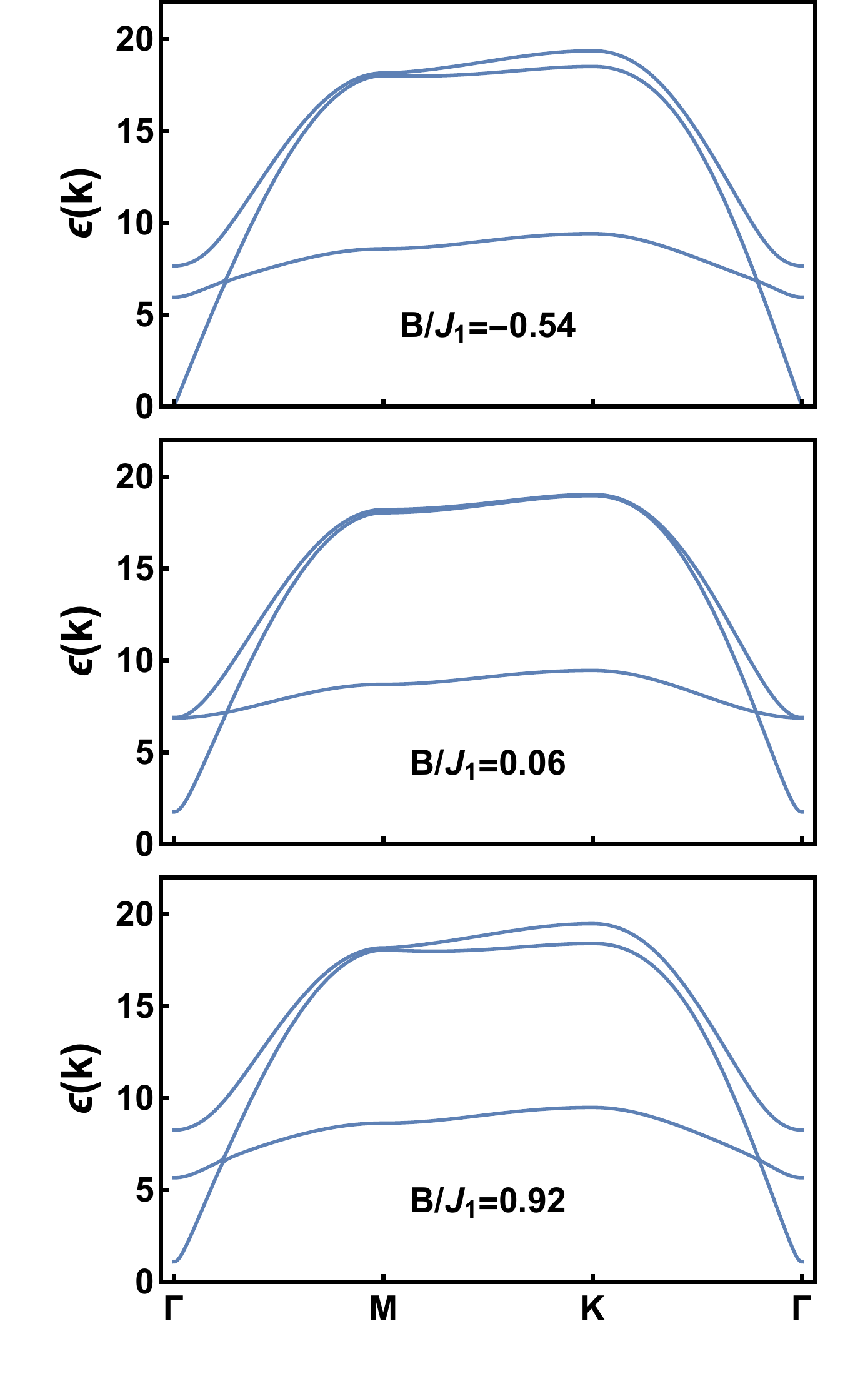}
		\caption{\label{fig:BandEvolutionB}Evolution of the magnon spectrum as the magnetic field strength $B$ is varied.}
	\end{center}
\end{figure}
\begin{figure}[tph!]
	\begin{center}
		\includegraphics[width=\columnwidth]{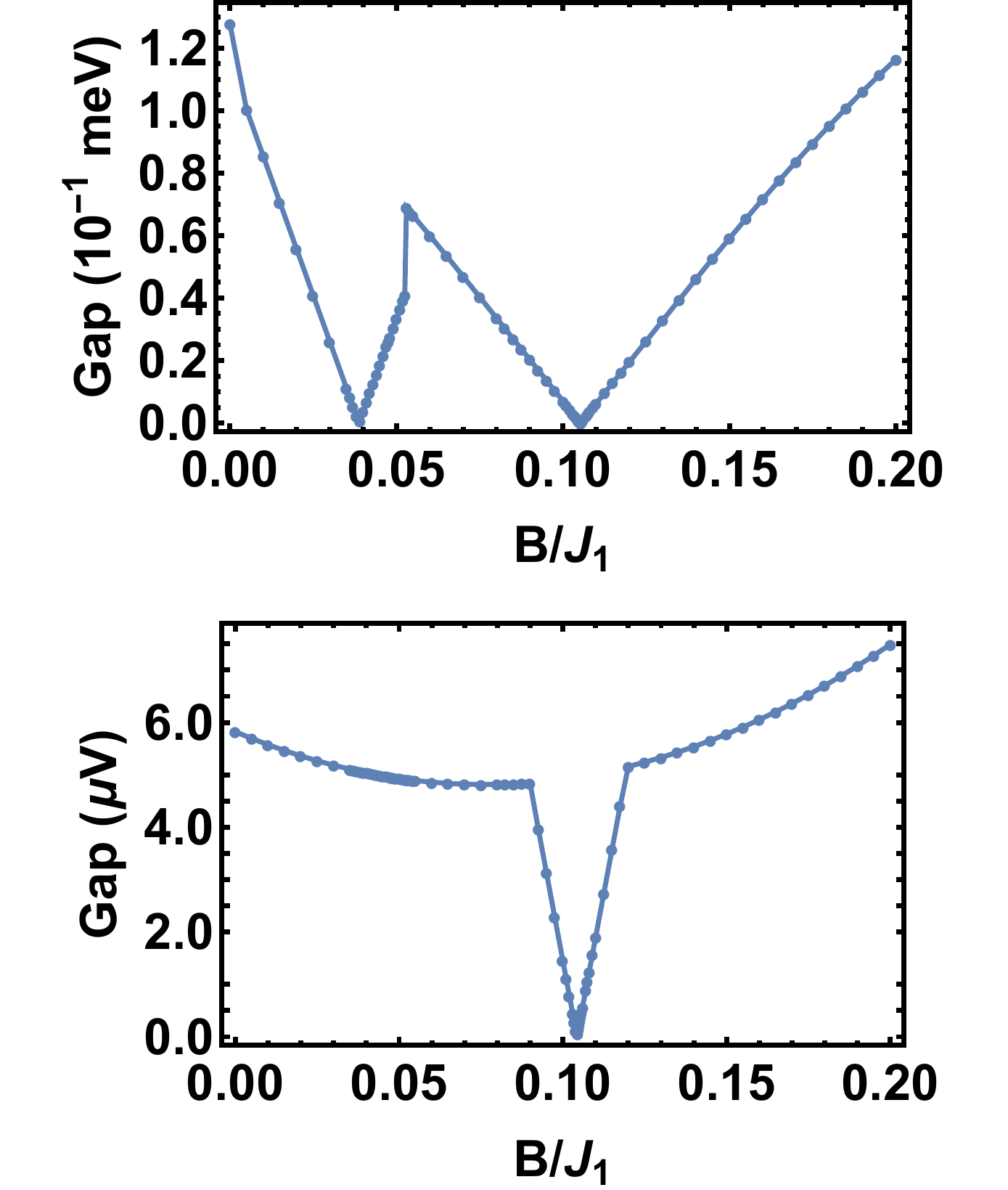}
		\caption{\label{fig:GapEvolutionB}Upper panel: Evolution of the upper gap as the magnetic field strength $B$ is varied. Lower panel: Evolution of the lower gap -- the avoided crossing.}
	\end{center}
\end{figure}

In Fig.~\ref{fig:BandEvolutionB} we plot the evolution of the spin wave spectrum for selected values of $B$. Near the topological transition the changes are relatively small, and hard to observe on this level of detail. We therefore plot the upper and lower gaps in Fig.~\ref{fig:GapEvolutionB}. The upper gap closes twice, at $B/J_1\approx 0.039$, and $B/J_1\approx 0.104$. The lower band closes only once, at $B/J_1\approx 0.105$. Reaching numerical convergence of the Chern numbers is challenging inbetween the two gap closings, but it does appear that the set of Chern numbers (from lowest to highest band) goes from $(-3,+1,+2)$ for $B<0.039J_1$, to $(-3,+3,0)$ for $0.039 \lesssim B/J_1 \lesssim 0.105$, to $(+3,-1,-2)$ for $B>0.105J_1$. The sign change of the thermal conductivity observed in Fig.~\ref{BScalingMatan} near $B/J_1=0.06$ can be related to the reversal in Chern numbers across this range of field strengths, but as this discussion shows, this reversal may have a substructure of consecutive topological transitions.

\begin{figure*}[tbp]
	\begin{center}
	\includegraphics[width=\textwidth]{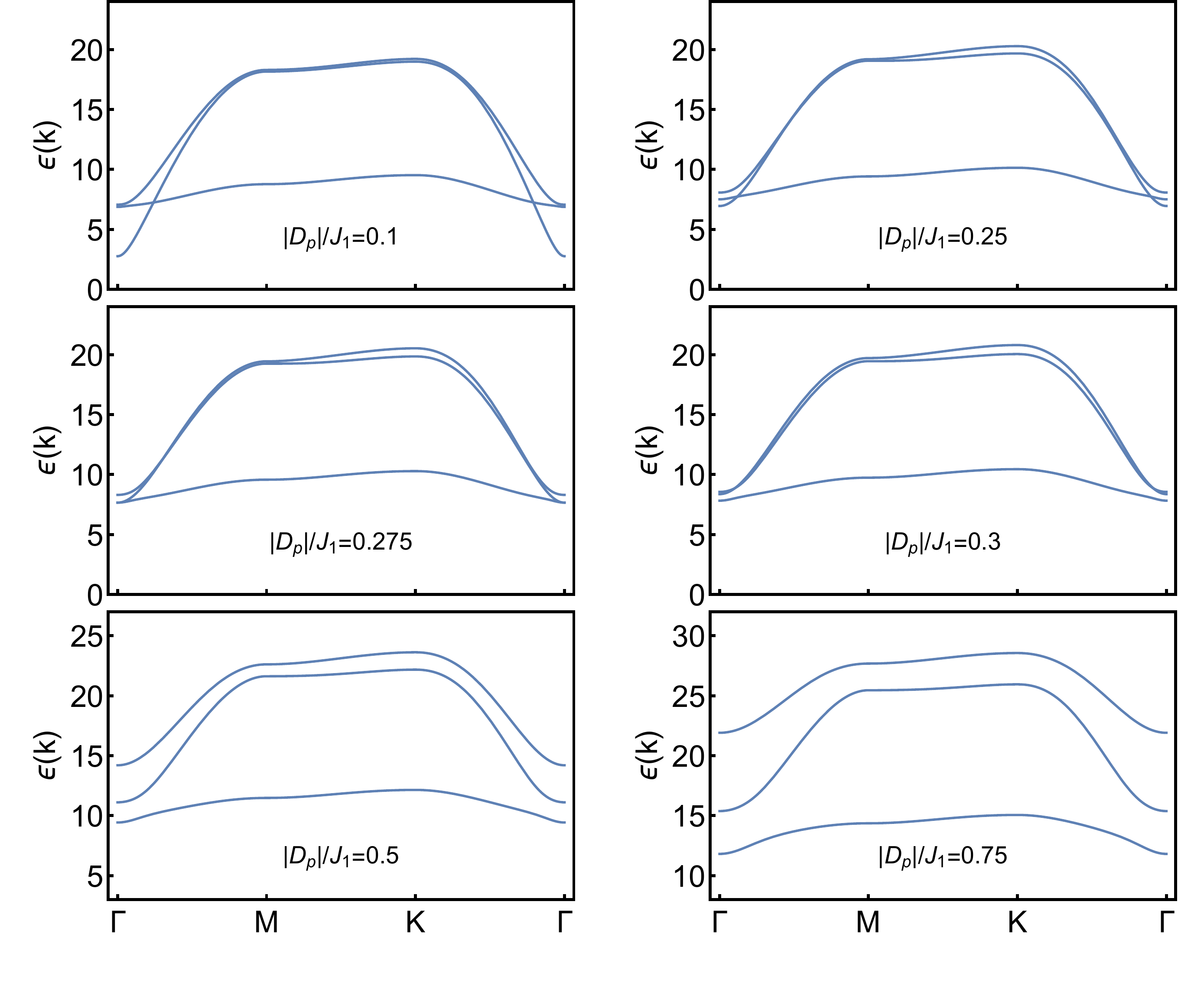}
	\caption{\label{fig:BandEvolution}Evolution of the magnon spectrum as the in-plane DMI strength $D_p$ is changed.}
	\end{center}
\end{figure*}
In Fig.~\ref{fig:BandEvolution} we plot the evolution of the magnon bands for different values of $D_p$. It is instructive to focus on the $\Gamma$ point. As $D_p$ is increased, the lowest band moves up and trades places with the middle band near the topological transition. Eventually the bands begin to separate and move to higher energies, explaining the downturn in Fig.~\ref{DpScalingMatan}.

\bibliographystyle{apsrev4-1}

%

\end{document}